\documentclass
[aps,prb,showpacs,twocolumn,amsmath,amssymb,a4paper,superscriptaddress]
{revtex4}
\usepackage{graphicx}
\usepackage{here}
\begin{document}

\title{Hard x-ray photoemission study of 
LaAlO$_3$/LaVO$_3$ multilayers}

\author{H.~Wadati}
\email{wadati@wyvern.phys.s.u-tokyo.ac.jp}
\homepage{http://www.geocities.jp/qxbqd097/index2.htm}
\altaffiliation[Present address: ]
{Department of Physics and Astronomy, 
University of British Columbia, Vancouver, 
British Columbia V6T-1Z1, Canada}
\affiliation{Department of Physics, 
University of Tokyo, 
Bunkyo-ku, Tokyo 113-0033, Japan}

\author{Y.~Hotta}
\affiliation{Department of Advanced Materials Science, 
University of Tokyo, Kashiwa, Chiba 277-8561, Japan}

\author{A.~Fujimori}
\affiliation{Department of Physics, 
University of Tokyo, 
Bunkyo-ku, Tokyo 113-0033, Japan}

\author{T.~Susaki}
\affiliation{Department of Advanced Materials Science, 
University of Tokyo, Kashiwa, Chiba 277-8561, Japan}

\author{H.~Y.~Hwang}
\affiliation{Department of Advanced Materials Science, 
University of Tokyo, Kashiwa, Chiba 277-8561, Japan}
\affiliation{Japan Science and Technology Agency, Kawaguchi 332-0012,
Japan}

\author{Y.~Takata}
\affiliation{Soft X-ray Spectroscopy Laboratory, 
RIKEN/SPring-8, 1-1-1 Kouto, Sayo-cho, 
Sayo-gun, Hyogo 679-5148, Japan}

\author{K.~Horiba}
\affiliation{Soft X-ray Spectroscopy Laboratory, 
RIKEN/SPring-8, 1-1-1 Kouto, Sayo-cho, 
Sayo-gun, Hyogo 679-5148, Japan}

\author{M.~Matsunami}
\affiliation{Soft X-ray Spectroscopy Laboratory, 
RIKEN/SPring-8, 1-1-1 Kouto, Sayo-cho, 
Sayo-gun, Hyogo 679-5148, Japan}

\author{S.~Shin}
\affiliation{Soft X-ray Spectroscopy Laboratory, 
RIKEN/SPring-8, 1-1-1 Kouto, Sayo-cho, 
Sayo-gun, Hyogo 679-5148, Japan}
\affiliation{Institute for Solid State Physics, 
University of Tokyo, Kashiwa, Chiba 277-8581, Japan}

\author{M.~Yabashi}
\affiliation{Coherent X-ray Optics Laboratory, 
RIKEN/SPring-8, 1-1-1 Kouto, Sayo-cho, Sayo-gun, 
Hyogo 679-5148, Japan}
\affiliation{JASRI/SPring-8, 1-1-1 Kouto, 
Sayo-cho, Sayo-gun, Hyogo 679-5198, Japan}

\author{K.~Tamasaku}
\affiliation{Coherent X-ray Optics Laboratory, 
RIKEN/SPring-8, 1-1-1 Kouto, Sayo-cho, Sayo-gun, 
Hyogo 679-5148, Japan}

\author{Y.~Nishino}
\affiliation{Coherent X-ray Optics Laboratory, 
RIKEN/SPring-8, 1-1-1 Kouto, Sayo-cho, Sayo-gun, 
Hyogo 679-5148, Japan}

\author{T.~Ishikawa}
\affiliation{Coherent X-ray Optics Laboratory, 
RIKEN/SPring-8, 1-1-1 Kouto, Sayo-cho, Sayo-gun, 
Hyogo 679-5148, Japan}
\affiliation{JASRI/SPring-8, 1-1-1 Kouto, 
Sayo-cho, Sayo-gun, Hyogo 679-5198, Japan}

\date{\today}
\begin{abstract}
We have studied the electronic structure 
of multilayers composed of a band 
insulator 
LaAlO$_3$ (LAO) and a Mott insulator LaVO$_3$ (LVO) 
by means of hard x-ray photoemission spectroscopy, which 
has a probing depth as large as $\sim 60\ \mbox{\AA}$. 
The Mott-Hubbard gap of LVO remained open at the 
interface, indicating that the interface is 
insulating unlike the 
LaTiO$_3$/SrTiO$_3$ multilayers. 
We found that the valence of V in LVO 
were partially converted from V$^{3+}$ 
to V$^{4+}$ only at the interface on the top side of 
the LVO layer and that the amount of V$^{4+}$ increased 
with LVO layer thickness.  
We suggest that the electronic reconstruction to 
eliminate the polarity catastrophe inherent 
in the polar heterostructure 
is the origin of 
the highly asymmetric valence change at the 
LVO/LAO interfaces. 
\end{abstract}
\maketitle
\section{Introduction}
The interfaces of hetero-junctions composed 
of transition-metal oxides have recently 
attracted great interest. For example, 
it has been suggested that 
the interface between a band insulator 
SrTiO$_3$ (STO) and a Mott 
insulator LaTiO$_3$ (LTO) shows metallic 
conductivity. \cite{HwangTi,shibuyaLTOSTO,okamotoLTOSTO} 
Recently, 
Takizawa {\it et al.} \cite{takizawaLTOSTO} 
measured photoemission spectra of this interface 
and observed a clear Fermi cut-off, indicating that 
an electronic reconstruction indeed  occurs 
at this interface. 
In the case of STO/LTO, 
electrons penetrate from the layers of 
the Mott insulator to the layers of the 
band insulator, resulting in the intermediate 
band filling and hence the metallic conductivity 
of the interfaces. 
It is therefore interesting to investigate 
how electrons behave if we 
confine electrons in the layers of 
the Mott insulator. 
In this paper, we investigate the 
electronic structure of multilayers 
consisting of 
a band insulator LaAlO$_3$ (LAO) and a Mott insulator 
LaVO$_3$ (LVO). 
LAO is a band insulator with a large band 
gap of about 5.6 eV. 
LVO is a Mott-Hubbard insulator 
with a band gap of about 1.0 eV. \cite{opt1} 
This material shows G-type orbital ordering 
and C-type spin ordering below the transition 
temperature $T_{OO}=T_{SO}=
143$ K. \cite{LVO2} 
From the previous results of 
photoemission and inverse photoemission spectroscopy, 
it was revealed that 
in the valence band there are O $2p$ bands at 
$4-8$ eV and V $3d$ bands (lower Hubbard bands; LHB) at 
$0-3$ eV and that above $E_F$ there are 
upper Hubbard bands (UHB) of V $3d$ origin 
separated by a band gap of 
about 1 eV from the LHB. \cite{Maiti} 
Since the bottom of the conduction band of LAO 
has predominantly La $5d$ character and its energy 
position is well above that of the LHB of LVO, \cite{LAO} 
the V $3d$ electrons are expected to be 
confined within the LVO 
layers as a ``quantum well'' 
and not to penetrate into the LAO layers, 
making 
this interface insulating unlike 
the LTO/STO case. \cite{HwangTi,shibuyaLTOSTO,
okamotoLTOSTO,takizawaLTOSTO} 
Recently, Hotta {\it et al.} \cite{Hotta} investigated 
the electronic structure of 1-5 unit cell thick layers 
of LVO embedded in LAO by means of soft x-ray (SX) 
photoemission spectroscopy. They found that the 
V $2p$ core-level spectra had both V$^{3+}$ and V$^{4+}$ 
components and that the V$^{4+}$ was localized 
in the topmost layer. However, due to the surface 
sensitivity of SX photoemission, 
the information 
about deeply buried interfaces in the multilayers 
is still lacking. Also, they used an 
unmonochromatized x-ray source, whose 
energy resolution was not sufficient for 
detailed studies of the valence band. 
In the present work, 
we have investigated the electronic structure 
of the LAO/LVO 
interfaces by means of 
hard x-ray (HX) 
photoemission spectroscopy 
($h\nu=7937$ eV) at SPring-8 BL29XU. 
HX photoemission spectroscopy is a 
bulk-sensitive experimental technique compared with 
ultraviolet and SX photoemission spectroscopy, 
and is very powerful for investigating buried 
interfaces in multilayers. 
From the valence-band spectra, we found that 
a Mott-Hubbard gap 
of LVO remained open at the interface, 
indicating the insulating nature of this interface. 
From the V $1s$ and $2p$ core-level spectra, 
the valence of V in LVO 
was found to be partially converted from V$^{3+}$ 
to V$^{4+}$ at the interface, confirming the 
previous study. \cite{Hotta} 
Furthermore, the amount of V$^{3+}$ was 
found to increase with LVO layer thickness. 
We attribute this valence change to 
the electronic 
reconstruction due to polarity of the layers. 

\section{Experiment}
\begin{figure}
\begin{center}
\includegraphics[width=9cm]{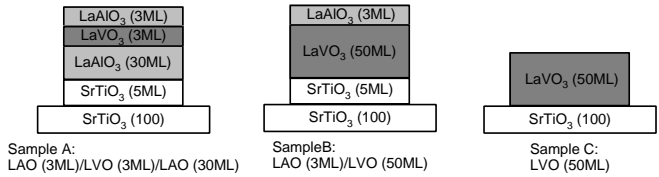}
\caption{Schematic view of the LaAlO$_3$/LaVO$_3$ 
multilayer samples. 
Sample A: LaAlO$_3$ (3ML)/LaVO$_3$ (50ML)/SrTiO$_3$. 
Sample B: LaAlO$_3$ (3ML)/LaVO$_3$ (50ML)/LaAlO$_3$ (30ML)/SrTiO$_3$. 
Sample C: LaVO$_3$ (50ML)/SrTiO$_3$.}
\label{fig1}
\end{center}
\end{figure}

The LAO/LVO multilayer thin films were fabricated 
on TiO$_2$-terminated STO$(001)$ substrates \cite{Kawasaki} 
using the pulsed laser deposition (PLD) 
technique. An infrared heating system was used for 
heating the substrates. The films were grown 
on the substrates at an oxygen pressure 
of $10^{-6}$ Torr using a KrF excimer laser 
($\lambda = 248$ nm) operating at 4 Hz. 
The laser fluency to ablate 
LaVO$_4$ polycrystalline and LAO single crystal targets 
was $\sim 2.5$ J/cm$^2$. The film 
growth was monitored using real-time 
reflection high-energy electron 
diffraction (RHEED). 
Schematic views of the fabricated thin films 
are shown in Fig.~\ref{fig1}. 
Sample A consisted of 3ML LVO capped with 
3ML LAO. Below the 3ML LVO, 
30ML LAO was grown, making LVO 
sandwiched by LAO. 
Sample B consisted of 50ML LVO 
capped with 3ML LAO. 
Sample C was 50ML LVO without LAO 
capping layers. 
Details of the fabrication and 
characterization of the films were 
described elsewhere. \cite{HottaLVO} 
The characterization of the electronic 
structure of uncapped LaVO$_3$ thin films 
by x-ray photoemission spectroscopy will be 
described elsewhere. \cite{WadatiuncapLVO} 
HX photoemission 
experiments were performed at an 
undulator beamline, BL29XU, of SPring-8. 
The experimental details are described in 
Refs.~\onlinecite{TamasakuHardX,IshikawaHardX,HardX}. The total 
energy resolution was set to about 180 meV. 
All the spectra were measured 
at room temperature. 
The Fermi level ($E_F$) position was determined by
measuring gold spectra.

\section{Results and discussion}
\begin{figure}
\begin{center}
\includegraphics[width=9cm]{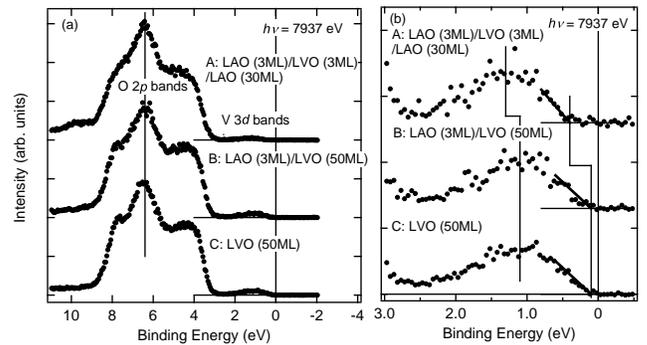}
\caption{Valence-band photoemission 
spectra of the LaAlO$_3$/LaVO$_3$ 
multilayer samples. 
(a) Valence-band spectra over a wide energy range. 
(b) V $3d$ band region.}
\label{fig2}
\end{center}
\end{figure}

Figure \ref{fig2} 
shows the valence-band photoemission 
spectra of the LAO/LVO 
multilayer samples.
Figure \ref{fig2} (a) shows the 
entire valence-band region. 
Compared with  
the previous photoemission results, \cite{Maiti} 
structures from 9 to 3 eV are assigned to the 
O $2p$ dominant bands, 
and emission from 3 eV to $E_F$ 
to the V $3d$ bands. 
The energy positions 
of the O $2p$ bands were almost the same in these 
three samples, indicating that the band bending 
effect at the interface of 
LAO and LVO was negligible. 
Figure \ref{fig2} (b) shows an 
enlarged plot of the spectra 
in the V $3d$-band region. 
A Mott-Hubbard gap of LVO remained open 
at the interface between LAO and 
LVO, indicating that this interface is 
insulating unlike the STO/LTO 
interfaces. \cite{HwangTi,shibuyaLTOSTO,
okamotoLTOSTO,takizawaLTOSTO} 
The line shapes of the V $3d$ bands were 
almost the same in these three samples, 
except for the energy shift in sample A. 
We estimated the value of the band gap from 
the linear extrapolation of the rising part of 
the peak as shown in Fig.~\ref{fig2} (b). 
The gap size of sample B was 
almost the same ($\sim 100$ meV) as that of 
sample C, while that of sample A was much larger 
($\sim 400$ meV) 
due to the energy shift of the V $3d$ bands. 
The origin of the enhanced energy gap is 
unclear at present, but an increase of 
the on-site Coulomb repulsion $U$ in the 
thin LVO layers compared to the thick LVO 
layers or bulk LVO due to a decrease of 
dielectric screening may explain the 
experimental observation. 

\begin{figure}
\begin{center}
\includegraphics[width=7cm]{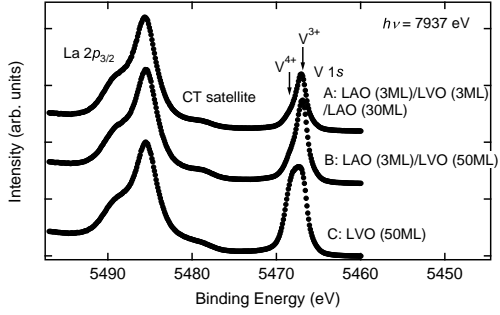}
\caption{V $1s$ core-level 
photoemission spectra of the LaAlO$_3$/LaVO$_3$ 
multilayer samples.}
\label{fig3}
\end{center}
\end{figure}

Figure \ref{fig3} 
shows the V $1s$ core-level photoemission spectra of the 
LAO/LVO multilayer samples. The 
V $1s$ spectra had a main peak at 
5467 eV and a satellite structure 
at 5478 eV. 
The main peaks were not simple symmetric peaks but 
exhibited complex line shapes. We therefore consider 
that the main peaks 
consisted of V$^{3+}$ and V$^{4+}$ 
components. In sample C, there is a considerable 
amount of V$^{4+}$ probably due to the oxidation of the 
surface of the uncapped LVO. A 
satellite structure has also been observed in the 
V $1s$ spectrum of V$_2$O$_3$ \cite{kamakura} 
and interpreted as a charge transfer (CT) 
satellites arising from 
the $1s^13d^3\underline{L}$ final state, 
where $\underline{L}$ denotes a hole in the O $2p$ band. 
Screening-derived peaks at the lower-binding-energy side 
of V $1s$, 
which have been observed in the metallic 
phase of V$_{2-x}$Cr$_x$O$_3$, \cite{kamakura,taguchi} 
were not observed in the present samples, 
again indicating the 
insulating nature of these interfaces. 

\begin{figure}
\begin{center}
\includegraphics[width=9cm]{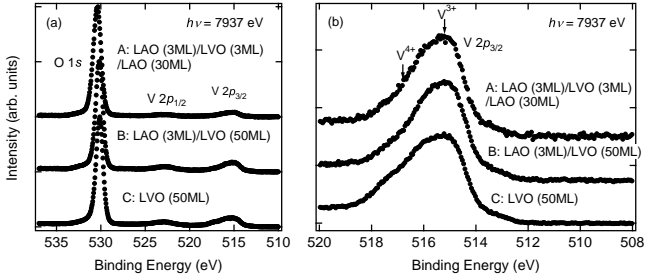}
\caption{
O $1s$ and V $2p$ core-level photoemission 
spectra of the LaAlO$_3$/LaVO$_3$ 
multilayer samples. 
(a) shows wide energy region and 
(b) is an enlarged plot of the 
V $2p_{3/2}$ spectra.}
\label{fig4}
\end{center}
\end{figure}

Figure \ref{fig4} 
shows the O $1s$ and V $2p$ core-level 
photoemission spectra of the 
LAO/LVO multilayer samples. 
The O $1s$ spectra consisted of single 
peaks without surface contamination signal 
on the higher-binding-energy side, 
indicating the bulk sensitivity of 
HX photoemission spectroscopy. 
The energy position of the O $1s$ peak 
of sample A, 
whose LVO layer thickness was 
only 3 ML, 
was different from those of 
the rest because LAO and LVO 
have different energy positions of the 
O $1s$ core levels.  
Figure \ref{fig4} (b) shows 
an enlarged plot of the 
V $2p_{3/2}$ spectra. Here again, 
the V $2p_{3/2}$ photoemission spectra 
showed complex line shapes consisting of 
V$^{3+}$ and V$^{4+}$ components, 
and 
no screening-derived peaks 
on the lower-binding-energy side 
of V $2p_{3/2}$ were observed. 
The line shapes of the V $2p_{3/2}$ spectra 
were very similar for samples A and B. 
The amount of 
V$^{4+}$ was larger in sample C, consistent with the case of 
V $1s$ and again shows the effect of the oxidation of the 
uncapped LVO.

\begin{figure}
\begin{center}
\includegraphics[width=9cm]{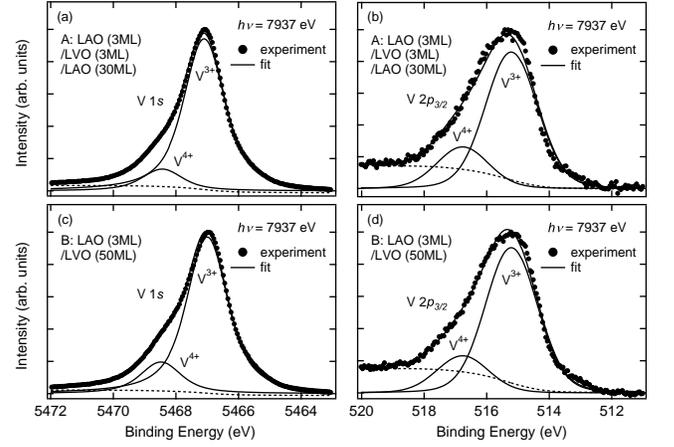}
\caption{Fitting results for the 
V $1s$ and $2p_{3/2}$ core-level spectra. 
(a) V $1s$ core level of sample A (LaVO$_3$ 3ML), 
(b) V $2p_{3/2}$ core level of sample A (LaVO$_3$ 3ML), 
(c) V $1s$ core level of sample B (LaVO$_3$ 50ML), 
(d) V $2p_{3/2}$ core level of sample B (LaVO$_3$ 50ML).}
\label{fig5}
\end{center}
\end{figure}

\begin{figure}
\begin{center}
\includegraphics[width=9cm]{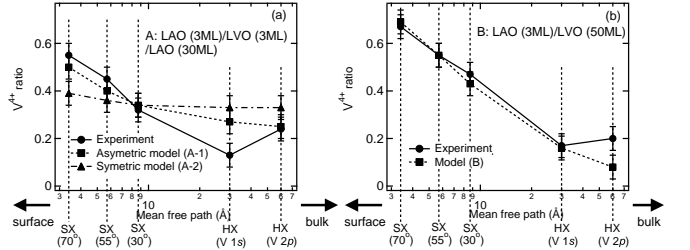}
\caption{Ratio of V$^{4+}$ or V$^{4+}+$ V$^{5+}$ 
determined under 
various experimental conditions using 
hard x-rays and soft x-rays. \cite{Hotta} 
(a) Sample A (3ML LaVO$_3$), 
(b) Sample B (50ML LaVO$_3$). Here, 
SX is a result of soft x-ray photoemission, and 
HX is of hard x-ray photoemission. 
In the case of SX, 
the values 
in the parenthesis denote the values of $\theta_e$.}
\label{fig6}
\end{center}
\end{figure}

We have fitted the core-level spectra 
of samples A and B to a Gaussian convoluted 
with a Lorentzian to estimate the amount of V$^{3+}$, 
V$^{4+}$ and V$^{5+}$ at the interface 
following the procedure of Ref.~\onlinecite{Hotta}. 
Figure \ref{fig5} shows the fitting results 
of the V $1s$ and V $2p_{3/2}$ 
core-level spectra. 
Here, the spectra were decomposed 
into the V$^{3+}$ and V$^{4+}$ components, and 
the V$^{5+}$ component was not necessary. 
The full width at half maximum (FWHM) 
of the Lorentzian has been 
fixed to 1.01 eV 
for V $1s$ and to 0.24 eV for V $2p_{3/2}$ 
according to Ref.~\onlinecite{width}. 
The FWHM of the Gaussian has been chosen 
0.90 eV for 
V $1s$ and 1.87 eV for V ${2p}_{3/2}$, 
reflecting 
the larger multiplet splitting 
for V $2p$ than for V $1s$. 
In Fig.~\ref{fig6}, we summarize 
the ratio of the V$^{3+}$ component 
thus estimated, together with 
the results of 
the emission angle ($\theta_e$) 
dependence of the V $2p$ core-level 
SX photoemission spectra 
measured using a laboratory 
SX source. \cite{Hotta} 

In order to interpret those results qualitatively, 
first we have to know the probing depth of 
photoemission spectroscopy under 
various measurement conditions. 
From the kinetic energies of photoelectrons, 
the mean free paths of the respective measurements 
are obtained as described in 
Ref.~\onlinecite{Tanuma}.
\cite{foot} 
When we measure V $2p_{3/2}$ spectra with 
the Mg K$\alpha$ line ($h\nu = 1253.6$ eV), 
the kinetic energy of photoelectrons is 
about 700 eV, and the mean free path is 
estimated to be about 10 $\mbox{\AA}$. 
Likewise, we also estimate the mean free path 
in the HX case. The values are summarized in 
Table~\ref{tab1}. 
In the SX case, these values are $10\cos\theta_e$ 
$\mbox{\AA}$. 
One can obtain the most surface-sensitive spectra 
under the condition of SX with $\theta_e=70^o$ 
[denoted by SX(70$^o$)] and 
the most bulk-sensitive spectra 
for HX measurements of the V $2p_{3/2}$ core level 
[denoted by HX(V $2p$)]. 
From Fig.~\ref{fig6} and Table~\ref{tab1}, 
one observes a larger amount of V$^{4+}$ components 
under more surface-sensitive conditions. 
These results demonstrate that the valence of V in LVO 
is partially converted from V$^{3+}$ 
to V$^{4+}$ at the interface. 

\begin{table}
\begin{center}
\caption{Mean free path of photoelectrons 
(in units of $\mbox{\AA}$)}
\begin{tabular}{|c|c|c|c|c|}
\hline
SX & SX & SX & HX & HX\\
(70$^{\circ}$) & (55$^{\circ}$) & 
(30$^{\circ}$) & (V $1s$) & (V $2p$)\\
\hline
3.4 & 5.7 & 8.7 & 30 & 60 \\
\hline
\end{tabular}
\label{tab1}
\end{center}
\end{table}
 
\begin{figure}
\begin{center}
\includegraphics[width=9cm]{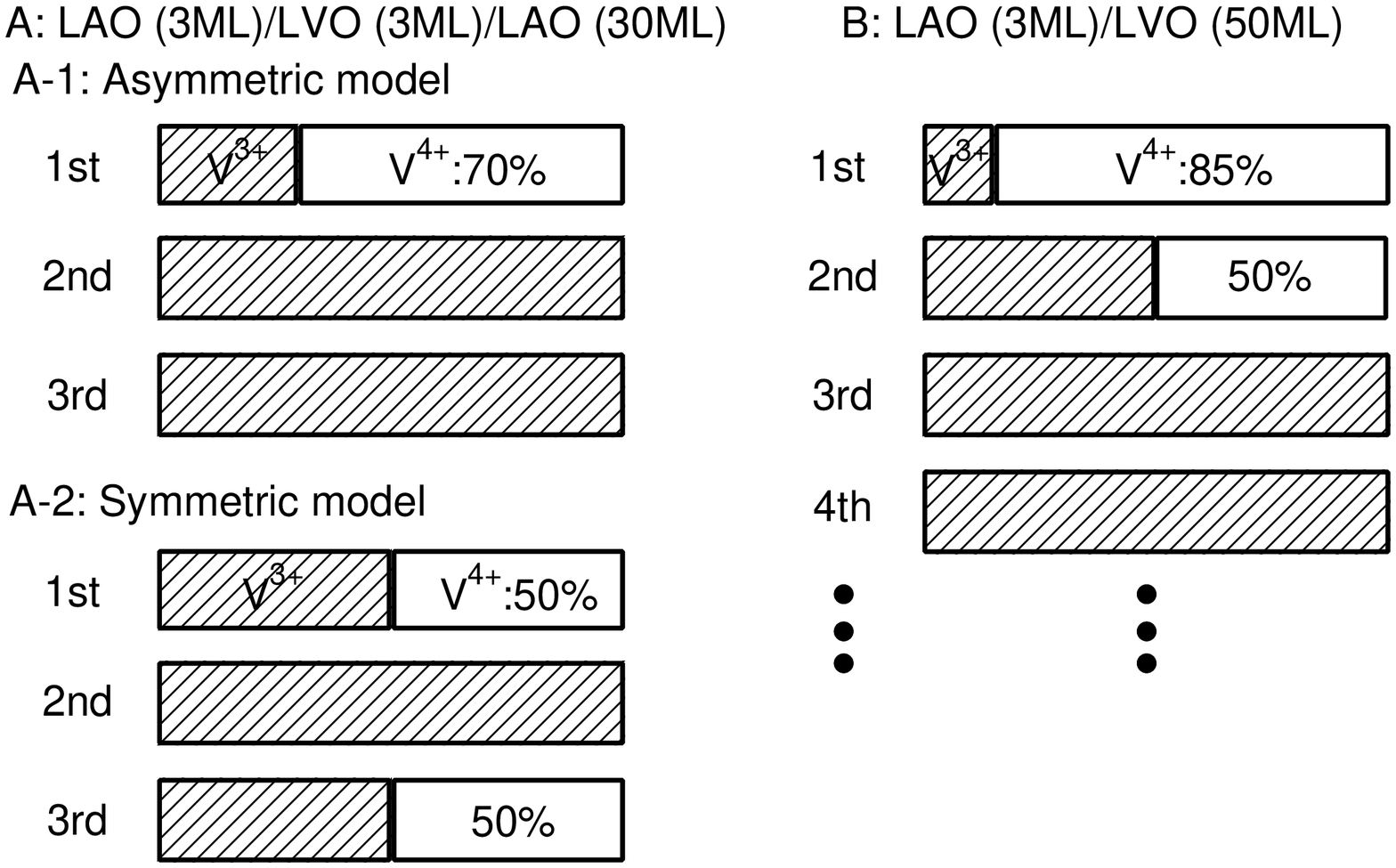}
\caption{Models for the V valence 
distributions in the LaAlO$_3$/LaVO$_3$ 
multilayer samples. 
A: LaVO$_3$ 3ML. A-1 is an asymmetric model, whereas 
A-2 is a symmetric model. 
B: LaVO$_3$ 50ML.}
\label{fig7}
\end{center}
\end{figure}

In order to reproduce the present experimental 
result and the result reported in Ref.~\onlinecite{Hotta} 
(shown in Fig.~\ref{fig6}), 
we propose a model of the V valence distribution 
at the interface as shown in Fig.~\ref{fig7}. 
For sample A, we consider two models, that is, 
an asymmetric model and a symmetric model. 
In the asymmetric model (A-1), 
no symmetry is assumed between 
the first and the third layers. 
As shown in Fig.~\ref{fig6}, 
the best fit result was obtained 
for the valence distribution that 
70 \% of the first layer is V$^{4+}$ 
and there are no V$^{4+}$ in the 
second and third layers, assuming 
the above-mentioned mean free paths in 
Table~\ref{tab1} and exponential decrease of 
the number of photoelectrons. 
In the symmetric model (A-2), 
it is assumed that 
the electronic structures are symmetric 
between the first and the third layers. 
The best fit was obtained when 
50 \% of the first and third layers 
are V$^{4+}$. 
In sample B, a model (B) where 
85 \% of the first layer and 
50 \% of the second layer are 
V$^{4+}$ best reproduced the 
experimental result. 
As shown in Fig.~\ref{fig6}, 
for the 3ML case, the model (A-2) did not reproduce 
the experimental results well compared to (A-1), 
which demonstrates that 
the valence distribution of V was 
highly asymmetric at these interfaces. 

The origin of this highly asymmetric valence 
change from V$^{3+}$ 
to V$^{4+}$ at the interfaces 
can be interpreted in two ways. 
One possible scenario is a simple chemical 
effect during the fabrication process 
of the PLD technique. The topmost LVO layer 
spends a longer time 
before the next deposition of LAO 
than the rest LVO 
layers, and therefore, oxidation process may 
easily proceed at the topmost layer. 
In this scenario, if we make 
samples under different experimental conditions, 
especially under different oxygen pressures, 
the amount of V$^{4+}$ at the interface may 
change greatly. 
In the other scenario, we 
consider that the polarity of the LAO/LVO 
multilayers plays an essential role. 
In the present samples, 
both the LAO and LVO layers are polar, and 
do not consist of charge neutral layers, 
that is, 
they consist of alternating stack of 
LaO$^+$ and AlO$_2^-$ or VO$_2^-$ layers. 
As recently discussed 
by Nakagawa {\it et al.}, \cite{nakagawa} 
electronic reconstruction occurs during the 
fabrication of the polar layers 
in order to prevent the divergence of 
Madelung potential, i.e., 
so-called polar catastrophe. \cite{catas} 
We consider that the electronic 
reconstruction occurs in the present samples, 
and that the valence change of V 
at the interface is a result of 
this reconstruction. 
This effect explains 
0.5 ML of V$^{4+}$, but we cannot 
explain the total amount of V$^{4+}$ exceeding 
0.5 ML, 
and we must also consider some 
chemical effects that V atoms are relatively 
easily oxidized at the topmost layer. 
Similar studies on samples with 
different termination layers 
will be necessary to test this scenario. 
Recently, 
Huijben {\it et al.} \cite{huijben} studied 
STO/LAO multilayers 
and found a critical thickness of LAO and STO, below 
which a decrease of the interface conductivity and 
carrier density occurs. Therefore, 
changing the numbers of LAO capping layers may also 
change the valence of V at the interface. 
Further systematic studies, including other systems like 
LTO/STO \cite{HwangTi,shibuyaLTOSTO,okamotoLTOSTO,takizawaLTOSTO} 
and LAO/STO 
\cite{nakagawa,huijben,Hwang2}, will reveal 
the origin of the valence change at the interface. 

\section{conclusion}
We have investigated the electronic structure 
of the multilayers composed of a band 
insulator 
LaAlO$_3$ and a Mott insulator LaVO$_3$ (LVO) 
by means of HX photoemission spectroscopy. 
The Mott-Hubbard gap of LVO remained open at the 
interface, indicating that the interface is 
insulating and the delocalization of $3d$ 
electrons does not occur 
unlike the 
LaTiO$_3$/SrTiO$_3$ multilayers. 
From the V $1s$ and $2p$ core-level photoemission 
intensities, we found that the valence of V in LVO 
was partially converted from V$^{3+}$ 
to V$^{4+}$ at the interface only on the top side of 
the LVO layer and that the amount of V$^{4+}$ increased 
with LVO layer thickness. 
We constructed a model for the V valence redistribution 
in order to explain the experimental result and found 
that the V$^{4+}$ is preferentially distributed on 
the top of the LVO layers. 
We suggest that the electronic reconstruction to 
eliminate polar catastrophe may be the origin of 
the highly asymmetric valence change at the interfaces. 
\section{acknowledgments}
The HX photoemission experiments reported here 
have benefited tremendously from the efforts 
of Dr.~D.~Miwa of the coherent x-ray optics 
laboratory RIKEN/SPring-8, Japan 
and we dedicate this work to him. 
This work was supported by a Grant-in-Aid 
for Scientific Research (A16204024) from 
the Japan Society for the Promotion of 
Science (JSPS) and a Grant-in-Aid 
for Scientific Research in Priority Areas 
``Invention of Anomalous Quantum Materials'' 
from the Ministry of Education, Culture, 
Sports, Science and Technology. 
H. W. acknowledges financial support from JSPS. 
Y. H. acknowledges support from QPEC,
Graduate School of Engineering, University of Tokyo. 
\bibliography{LVO1tex}
\end{document}